\documentclass[aps,prb,reprint,superscriptaddress,showpacs,citeautoscript, longbibliography]{revtex4-1}
\usepackage{amsmath,amssymb,bm}
\usepackage{graphicx}
\usepackage{epstopdf}
\usepackage{latexsym}
\usepackage{subfigure}
\usepackage{color}
\usepackage{hyperref}
\usepackage{xspace}

\newcommand{\hamilt}{{\widehat{\mathcal{H}}}}

\newcommand{\aver}[1]{\left\langle #1 \right\rangle}

\newcommand{\DIMPY}{(C$_{7}$H$_{10}$N)$_{2}$CuBr$_{4}$}

\begin{document}

\title{Scaling of temporal correlations in an attractive Tomonaga-Luttinger spin liquid}

\author{K. Yu. Povarov}
   \affiliation{Neutron Scattering and Magnetism, Laboratory for Solid State Physics, ETH Z\"{u}rich, Switzerland}

\author{D. Schmidiger}
    \affiliation{Neutron Scattering and Magnetism, Laboratory for Solid State Physics, ETH Z\"{u}rich, Switzerland}

\author{N. Reynolds}
   \affiliation{Neutron Scattering and Magnetism, Laboratory for Solid State Physics, ETH Z\"{u}rich, Switzerland}

\author{R. Bewley}
   \affiliation{ISIS Facility, Rutherford Appleton Laboratory, Chilton, Didcot, Oxon OX11 0QX, United Kingdom}

\author{A. Zheludev}
   \affiliation{Neutron Scattering and Magnetism, Laboratory for Solid State Physics, ETH Z\"{u}rich, Switzerland}

\date{\today}

\begin{abstract}
We report temperature-dependent neutron scattering measurements of
the local dynamic structure factor in the quantum spin ladder
\DIMPY\ in a magnetic field $H=9$~T, in its gapless quantum-critical
phase. We show that the measured quantity has a scaling form
consistent with expectations for a Tomonaga-Luttinger liquid with
attraction. The measured Luttinger parameter $K\approx1.25$  and
scaling function are in excellent agreement with
density matrix renormalization group numerical calculations for the underlying spin Hamiltonian.
\end{abstract}

%75.10.Kt    Quantum spin liquids, valence bond phases and related phenomena
%75.40.Gb    Dynamic properties (dynamic susceptibility, spin waves, spin diffusion, dynamic scaling, etc.)
%75.10.Jm    Quantized spin models, including quantum spin frustration
%78.70.Nx    Neutron inelastic scattering


\maketitle

Landau's Fermi liquid theory is the basis of our understanding of
interacting fermions, be it metals or neutron stars
~[\onlinecite{Landau1957,Landau1958,LLcourse_IX,AbrikosovMetals}]. Landau's
main argument regarding the stability of quasiparticles breaks down
in low dimensions. Interestingly, in one dimension, interacting
fermions still bear a unified description, known as the
Tomonaga-Luttinger liquid (TLL)
~[\onlinecite{TLL_Tomonaga,TLL_Luttinger,TLL_LiebMatthis,Giamarchibook}]. All
low-energy properties of such fermions, including thermodynamics,
correlation functions, susceptibilities, etc.,  are predicted to be
{\it universal} in that any details of the interaction potential are
irrelevant. Instead, the interactions are characterized by a single
dimensionless quantity, known as the Luttinger parameter $K$. Free
fermions correspond to $K=1$, $K<1$ implies repulsive interactions,
and fermions with attraction have $K>1$.

Experimental validations of this astonishingly strong statement of
universality are of utmost importance. The most obvious model
systems are one-dimensional metals in charge-transfer salts
~[\onlinecite{Zwyck1997}], quantum wires~[\onlinecite{Yacoby1996}], and quantum Hall
effect edge states~[\onlinecite{Chang1996,Grayson1998,Grayson2007}].
However, TLLs with the most experimentally accessible correlation
functions are found in seemingly unlikely places, namely, in magnetic
insulators. Following Haldane's application of the TLL description
to $S=1/2$ spin chains~[\onlinecite{Haldane1980}], it was realized that a
great variety of one-dimensional quantum magnets fall into this
universality class. The benefit of such mapping is that spin
correlations can be directly probed with neutron scattering, nuclear magnetic resonance (NMR),
electron spin resonance (ESR) and other techniques. A huge success of this approach was
measurements of universal finite-temperature scaling laws for
correlation functions in Heisenberg $S=1/2$ chains
~[\onlinecite{Dender1997thesis,KCuF3scaling}]. Similar work was done on
non-TLL critical systems, such as gapless spin ladders with cyclic
exchange~[\onlinecite{Lake2010confinement}]. However, Heisenberg spin chains
are but a very particular case of a {\it repulsive} TLL, with
$K=1/2$~[\onlinecite{Giamarchibook,Schulz1986}]. Fermions (electrons) in
quantum wires and one-dimensional metals are usually also repulsive.
Very recently, it was shown that an {\it attractive} TLL, previously
only known in certain quantum Hall edge states~[\onlinecite{Grayson2007}],
can also be realized in magnetized $S=1/2$ antiferromagnetic (AF)
Heisenberg spin ladders~[\onlinecite{PositiveK,DIMPYthermodynamics}]. NMR
experiments on a prototypical ladder compound confirmed this result
~[\onlinecite{DIMPYnmr}]. However, those studies covered less than an order
of magnitude in temperature at a single measurement frequency, which
is insufficient to establish power law scaling and universality. In
contrast, in the present Rapid Communicatioin we use neutron spectroscopy to observe
power law scaling of local temporal correlations in the same
material over more than two decades in $\omega/T$. Specifically, we
find that the measured scaled local dynamic structure factor
$T^{1/2K-1}\mathcal{S}(\omega)$ is a {\it universal} function of
$\omega/T$ in wide temperature and energy ranges with $K>1$.

The spin Hamiltonian for the AF Heisenberg spin ladder is written as:
 \begin{equation}
 \hamilt=\sum_j\left[J_\parallel\mathbf{S}_{1,j}\mathbf{S}_{1,j+1}+ J_\parallel\mathbf{S}_{2,j}\mathbf{S}_{2,j+1}+ J_\perp\mathbf{S}_{1,j}\mathbf{S}_{2,j}\right],
 \end{equation}
where $J_\parallel>0$ and $J_\perp>0$ are exchange constants for the
ladder legs 1 and 2, and rungs, respectively. In zero
magnetic field,  the ground state is a non-magnetic singlet and the
excitation spectrum is gapped. However, applying a magnetic field
drives this gap to zero at some critical field $H_{c1}$ by virtue of the
Zeeman effect, and produces a continuum of gapless TLL states for
$H>H_{c1}$. The Luttinger parameter can be continuously tuned by
varying the field strength~[\onlinecite{Giamarchibook}]. For $H\rightarrow
H_{c1}$, $K\rightarrow 1$ in all cases. For $J_\parallel\ll
J_{\perp}$, the quasiparticles can be envisioned as single-rung
singlet-triplet excitations. These have hard-core like repulsion,
since no rung can be excited twice. As a result, $K<1$ for
$H>H_{c1}$~[\onlinecite{BouilottNegativeK}]. However, if the leg
interactions are dominant, $J_{\parallel}>J_{\perp}$, the
quasiparticles are more extended in nature. For this case, numerical
calculations show $K>1$ above the critical field
~[\onlinecite{DIMPYthermodynamics}].

\begin{figure}
\begin{center}
  \includegraphics[width=0.5\textwidth]{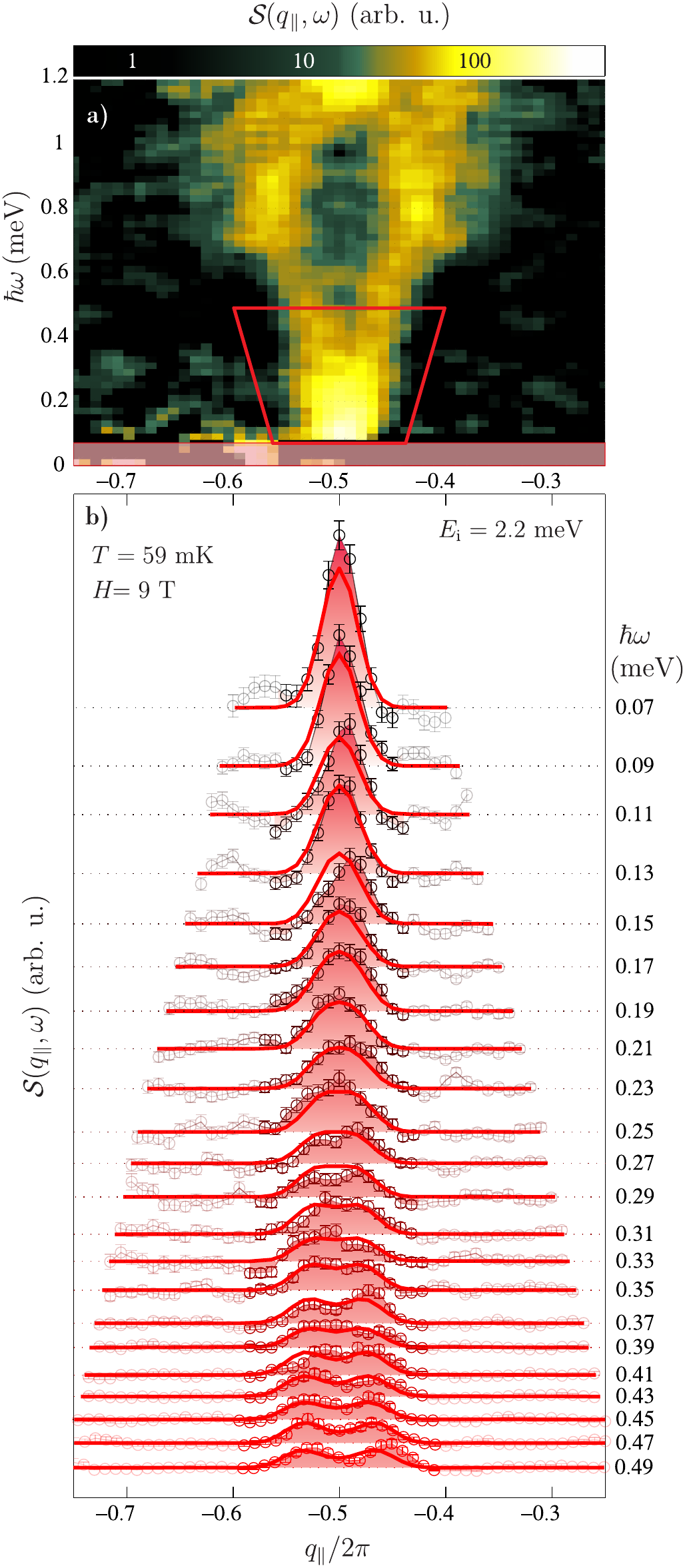}\\
  \caption{(Color online) Dynamic spin structure factor of DIMPY at $T=59$~mK, extracted from inelastic neutron intensities.  (a) False color map of $\mathcal{S}(q_{\parallel},\omega)$. The boundary shows the region in energy-momentum space,
  used in the analysis. (b) Constant energy cuts with the $q_{\parallel}$ integration range highlighted. All the data are background corrected, as described in the text. Solid lines correspond to the TLL theoretical prediction convoluted with the spectrometer resolution.}
  \label{FIG:LowT}
\end{center}
\end{figure}

\begin{figure} \begin{center}
  \includegraphics[width=0.5\textwidth]{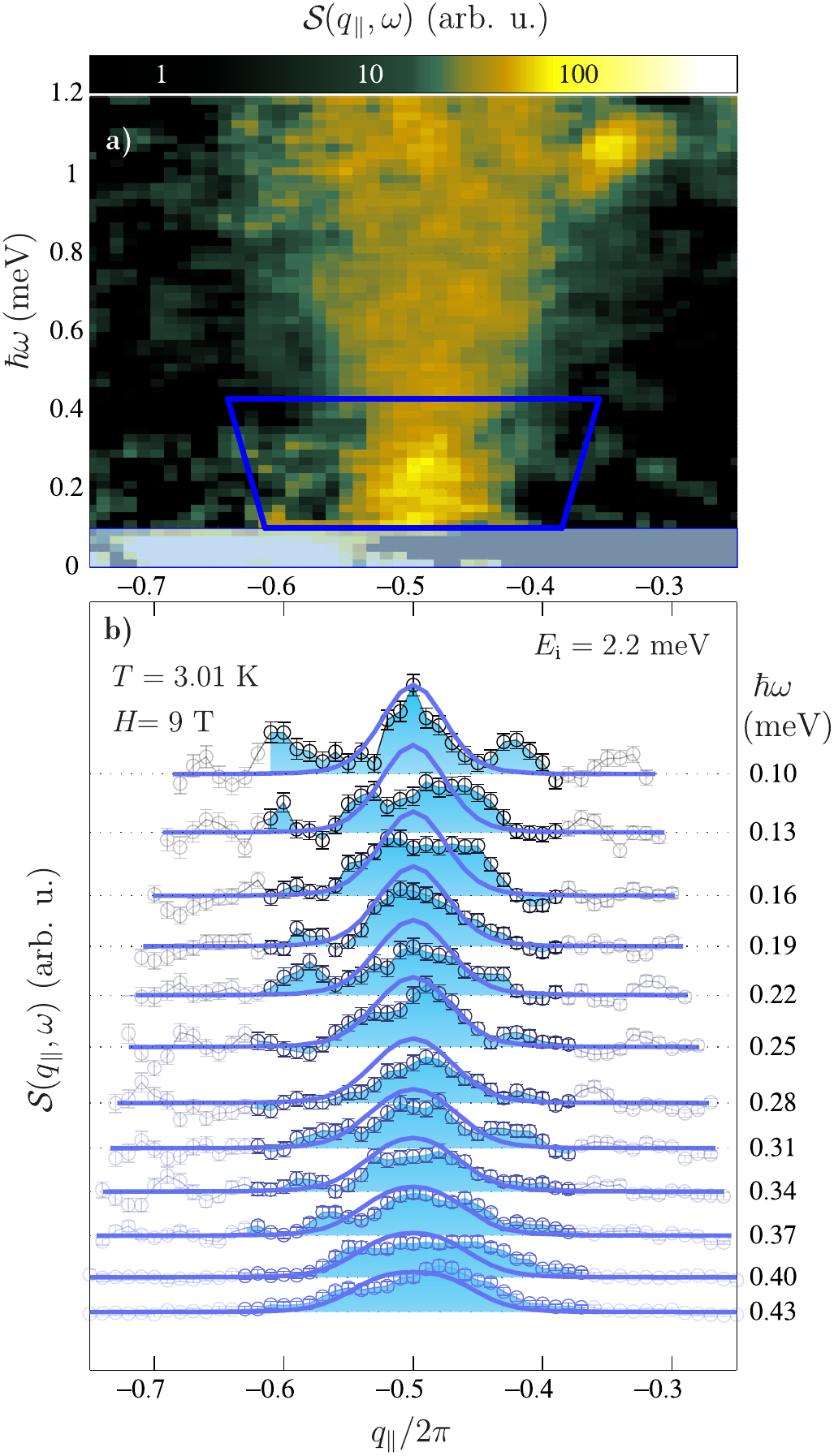}\\
  \caption{(Color online) As Fig.~\ref{FIG:LowT}, for $T=3.01$~K.}\label{FIG:HighT}
\end{center}
\end{figure}

Our target material, \DIMPY, bis(2,3-dimethylpyridinium) tetrabromocuprate or DIMPY for short, is arguably the best
known realization of the AF $S=1/2$ Heisenberg ladder model
~[\onlinecite{Hong2010,DIMPYmagnons,DIMPYthermodynamics,DIMPYprl}]. The
magnetic properties are due to Cu$^{2+}$ cations linked in ladder
structures by superexchange pathways via Br$^-$ anions
(for a schematic depiction of the crystal structure see Fig.~1 in Ref.~\onlinecite{DIMPYmagnons}). Organic ligands act as spacers between these
ladders in the monoclinic crystal structure, providing excellent
one-dimensionality. The spin Hamiltonian of DIMPY is well
established through a quantitative comparison of inelastic neutron
scattering spectra and thermodynamic data to  density
matrix renormalization group (DMRG) numerical calculations
~[\onlinecite{DIMPYmagnons,DIMPYprl}]. The two ladder exchange constants are
$J_{\parallel}=1.42$~meV and $J_{\perp}=0.82$~meV,
respectively. Interladder interactions are as small as $J'\simeq
6~\mu$eV~[\onlinecite{DIMPYthermodynamics}]. The ground state is a spin
singlet with a  gap $\Delta=0.33$~meV. In DIMPY, the spin gap closes at
$H_{c1}\simeq2.6$~T
~[\onlinecite{DIMPYthermodynamics,DIMPYnmr,DIMPYsymmetric}].

Measuring those temporal
correlations in DIMPY that are relevant to TLL physics is far from straightforward. Most of the observable spectral features are specific to spin ladder
physics and are not related to TLL dynamics~[\onlinecite{DIMPYsymmetric}].
TLL excitations are revealed only at the lowest energies, and are
barely discernible in previous measurements due to limited energy
resolution. Therefore, our present experiments, while using the same
sample \footnote{In all neutron experiments we used fully deuterated samples to avoid the strong incoherent scattering from H nuclei.}  and the LET neutron spectrometer~[\onlinecite{LETinstrument}] at ISIS
as in previous studies~[\onlinecite{DIMPYsymmetric}], took advantage of a
different instrument configuration with a lower neutron incident
energy $E_\mathrm{i}=2.2$~meV. This allowed us to achieve a calculated energy
resolution of $\delta E\sim20-30~\mu$eV, depending on energy
transfer.

Neutron experiments are used to determine the dynamic structure factors $\mathcal{S}^{\alpha\alpha}(\mathbf{Q},\omega)$ as a function of momentum transfer $\mathbf{Q}$ and energy transfer $\hbar \omega$. For each spin component $\alpha$, these quantities are a Fourier transform of the spin correlation functions of interest: $\mathcal{S}^{\alpha\alpha}(\mathbf{Q},\omega)=\int e^{-i\boldsymbol{(}(\mathbf{Q\cdot r})-\omega t\boldsymbol{)}}\aver{S^{\alpha}(0,0)S^{\alpha}(\mathbf{r},t)}d\mathbf{r}dt$.
In practice, one measures the neutron partial differential cross-section, which in our case is proportional to the sum of structure factors for spin components parallel and perpendicular to the applied field: $\mathcal{S}(\mathbf{Q},\omega)=\mathcal{S}^{zz}(\mathbf{Q},\omega)+\mathcal{S}^{\perp \perp}(\mathbf{Q},\omega)$~[\onlinecite{SquiresNeutron}].

The thus-defined structure factor $\mathcal{S}(\mathbf{Q},\omega)$,  measured at $H=9$~T and $T=59$~mK, is shown in Fig.~\ref{FIG:LowT}a. Here and below, it is plotted as a function of $q_{\parallel}=(\mathbf{Q\cdot a})$, where the $a$ crystallographic axis  is along the ladder direction. Previous DMRG calculations identified all the main components of the
excitation spectrum~[\onlinecite{DIMPYsymmetric}]. Thus, we know that the
prominent gapped incommensurate excitations seen at $H=9$~T at
around $\approx 0.8$~meV energy transfer are due to longitudinal
spin correlations $\mathcal{S}^{zz}(q_\parallel,\omega)$ and are not
TLL related. Since they can not be avoided in the experiment, our
analysis only considers the data below $\hbar\omega=0.5$~meV. Here it is due to correlations between transverse spin components,
so $\mathcal{S}(q_\parallel,\omega)\simeq\mathcal{S}^{\perp \perp}(q_\parallel,\omega)$. The latter is representative of TLL correlations, and is
the main focus of this study.
The scattering at these low energies is a \textsf{V}-shaped continuum visible
in Fig.~\ref{FIG:LowT}(a) around the commensurate point
$q_{\parallel}=-\pi$.
The slope of the lower bound of the
observed continuum, discernible as the separation between the two
separate peaks seen at higher energies in Fig.~\ref{FIG:LowT},
corresponds to the Fermi velocity of the TLL. Since $q_{\parallel}$
is dimensionless in our notation, for consistency the velocity is
measured in energy units, so that $\hbar\omega=uq_{\parallel}$ for a
linear dispersion relation. The experimental value
$u=1.98\pm0.02$~meV is in excellent agreement with the DMRG result
$u=1.91$~meV for DIMPY at $H=9$~T~[\onlinecite{DIMPYthermodynamics}].

\begin{figure} \begin{center}
  \includegraphics[width=0.5\textwidth]{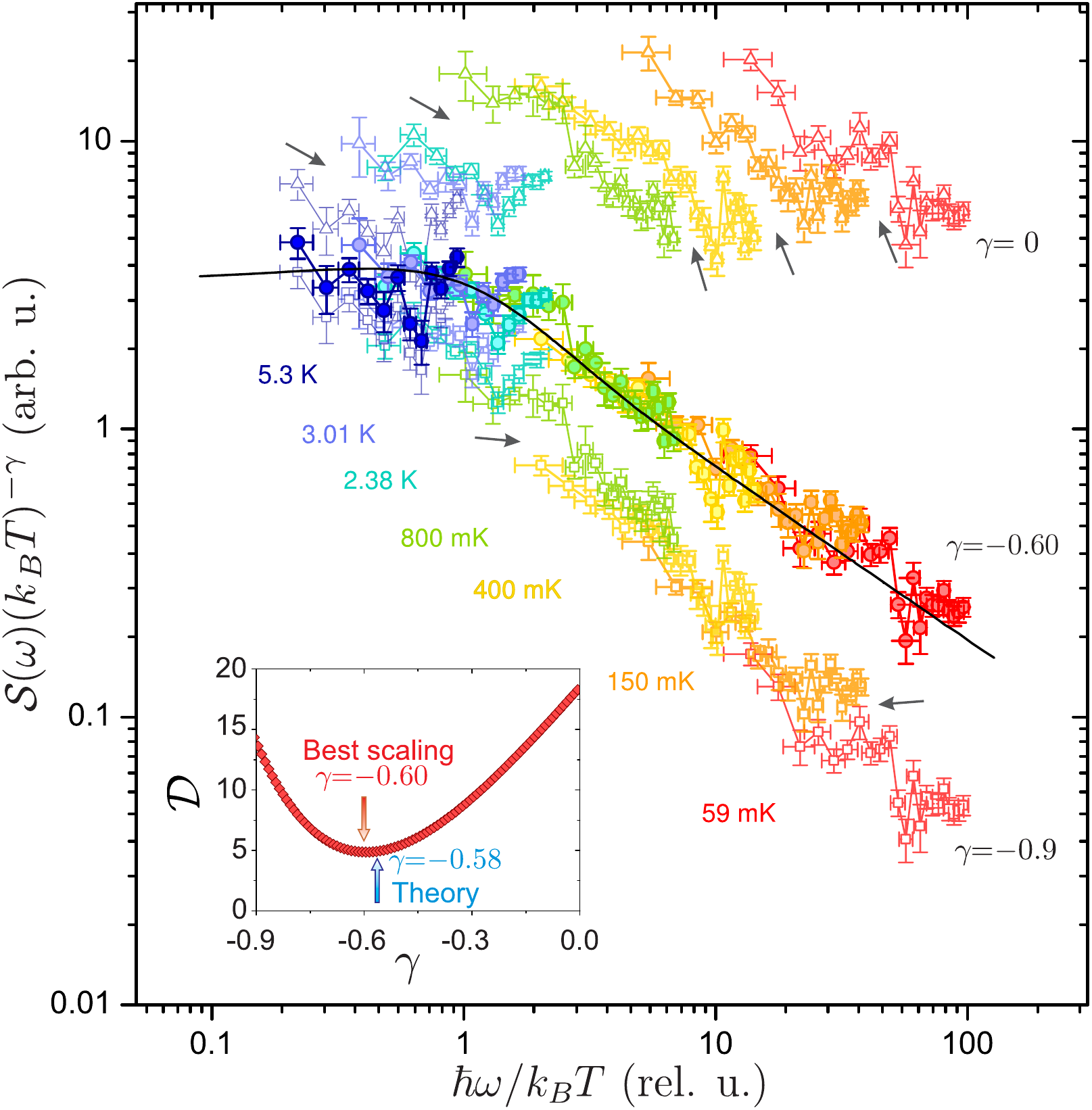}\\
  \caption{(Color) Local dynamic structure factor $\mathcal{S}(\omega)$ measured in DIMPY at $H=9$~T at several temperatures, plotted in the scaling representation with different scaling exponents $\gamma$. Arrows mark the apparent
  violations of scaling for non-optimal values of $\gamma$. The solid line
  is the exact TLL scaling function $F$ [Eq.~(\ref{EQ:Scaling})] with the Luttinger parameter $K=1.25$, corresponding to $\gamma=-0.6$.  Inset: $\gamma$ dependence of data overlap  function $\mathcal{D}$ [Eq.~(\ref{EQ:Deviation})].}\label{FIG:Cracked}
\end{center} \end{figure}

Having chosen a suitable measurement window, we still need to
ensure that the TLL mapping remains valid in that range. The essences
of the TLL model are (i) a linear dispersion relation for the
fermions and (ii) an infinitely deep Fermi sea. It is a valid
approximation only if the experimental energy transfers and
temperature do not probe the actual Fermi sea too deeply or extend
to non-linear dispersion regimes.  Looking at previous DMRG
calculations for $\mathcal{S}^{\perp \perp}(q_{\parallel},\omega)$ at a similar
field $H=7$~T, we conclude that in our case (i) holds to at least
1~meV energy transfer. At $H=9$~T the depth of the Fermi sea is
$\Delta_\mathrm{F}=g\mu_\mathrm{B}(H-H_{c1})\sim 0.77$~meV. Since we
have already limited ourselves to excitations below 0.5~meV, the TLL
model should be applicable with a good safety margin, as long as we
also limit the experimental temperatures to $T<5$~K$\sim
0.5$~meV$/k_{B}$.

Constant-energy cuts from two data sets collected at the base
temperature and at $T=3$~K are shown in Figs.~\ref{FIG:LowT}(b) and
\ref{FIG:HighT}(b), respectively. Here the background, measured in
zero applied field at the base temperature, as described in
Ref.~\onlinecite{DIMPYsymmetric}, was subtracted.
At $H=9$~T, the finite {\it momentum} resolution
of our experiments prevents us from discerning any internal
structure of the excitation continuum below 0.25~meV. As was done previously for $S=1/2$ chain systems
~[\onlinecite{Dender1997thesis,BaCuSiGeO2007,SrCuO2Pseudogap}], this problem
was entirely avoided by integrating the dynamic structure factor
over $q_{\parallel}$. The integration range was limited to the
trapezoidal areas indicated in Figs.~\ref{FIG:LowT}(a) and
\ref{FIG:HighT}(a), to avoid picking up additional noise. The
resulting momentum-integrated structure factor is the {\it local}
temporal transverse spin correlation function in frequency
representation $\mathcal{S}^{\perp \perp}(\omega)$.

We chose to analyze these data in two separate stages. First, we consider the temperature
dependence of the measured structure factor without referring to specific results for the TLL.
In fact, our only starting consideration is that the magnetized spin ladder is in a quantum-critical
state~[\onlinecite{Sachdev_1DAFMS}], and that $\mathcal{S}^{\perp \perp}(\omega)$ is precisely the critical correlation function. In this case, the only
relevant energy scale is temperature itself, so that the {\it shape} of $\mathcal{S}^{\perp \perp}(\omega)$ is a function of a
single variable of $\omega/T$. One can then expect the following scaling form for this quantity~[\onlinecite{SachdevBook}]:
\begin{equation}
\mathcal{S}^{\perp \perp}(\omega)=(k_{B}T)^{\gamma}F(\hbar\omega/k_{B}T).\label{EQ:Scaling}
\end{equation}
If so, with an appropriate choice of the scaling exponent $\gamma$,
we should observe a data collapse for measurements collected at
different temperatures. For illustration purposes,
Fig.~\ref{FIG:Cracked}a shows scaling plots of our data obtained
with three different exponents.  For $\gamma=0$ and $\gamma=-0.9$
the data collapse is visually less than perfect, with discontinuities indicated by arrows. For a more meaningful discussion, though, we introduced an {\it ad hoc} measure of the quality of the data overlap,
\begin{equation}
\mathcal{D}=\sum\limits_{n,i,j}\dfrac{\left(\sigma_{n,i}-\sigma_{n+1,j}\right)^2}{\left(\Delta\sigma_{n,i}+\Delta\sigma_{n+1,j}\right)^{2}}\frac{1}{\sum_{n,i,j}},\label{EQ:Deviation}
\end{equation}
where $\sigma_{n,i}=\mathcal{S}(\omega_{ni})|_{T=T_{n}}/(k_{B}T_{n})^{\gamma}$,
$i,j$ enumerate the nearest-neighbor pairs of points in data subsets $T_{n}$ and $T_{n+1}$,
and $\Delta\sigma_{n,i}$ is the corresponding error.
The inset of Fig.~\ref{FIG:Cracked} shows the $\gamma$ dependence of
$\mathcal{D}$. The thus defined data overlap  is optimized with
$\gamma=-0.60$, where $\mathcal{D}$ has a minimum, as indicated by the arrow in the inset of 
Fig.~\ref{FIG:Cracked}. The corresponding scaling plot of
$\mathcal{S}(\omega)$ is shown as solid symbols in the  main panel Fig.~\ref{FIG:Cracked}.
An excellent overlap between data sets collected at different
temperatures is apparent.

Having experimentally established the scaling form of the local dynamic structure
factor, we can compare the result to predictions of
TLL theory. In this model the dynamic structure factor scales with a
known scaling function and exponent
~[\onlinecite{Giamarchibook,SachdevBook}]:
\begin{equation}
    \begin{aligned}
&\mathcal{S}^{\perp \perp}(q_{\parallel},\omega)\propto  T^{1/2K-2} \times \\
& \times\Im\left\{ \left[ 1-\exp\left(-\frac{\hbar \omega}{k_{B}T}\right)\right]^{-1} \Phi\left(\frac{\hbar \omega}{k_{B}T},\frac{ u(q_{\parallel}-\pi)}{k_{B}T}\right)\right\},\\
&\Phi(x,y)=\frac{\Gamma(\frac{1}{8K}-i\frac{x-y}{4\pi})}{\Gamma(1-\frac{1}{8K}-i\frac{x-y}{4\pi})}\frac{\Gamma(\frac{1}{8K}-i\frac{x+y}{4\pi})}{\Gamma(1-\frac{1}{8K}-i\frac{x+y}{4\pi})}.\label{EQ:Sfactor}
    \end{aligned}
\end{equation}

Integrating over momentum and comparing this to
Eq.~\ref{EQ:Scaling} yields $\gamma=\frac{1}{2K}-1$. From the
experimentally determined scaling exponent $\gamma$ we then get
$K\approx 1.25$. Thus, we are dealing with one-dimensional fermions
{\it with attraction}: $K>1$. The comparison with TLL theory can be
taken further. By numerically integrating the function $\Phi$ in
Eq.~\ref{EQ:Sfactor}, using $K=1.25$ and an arbitrary
overall scale factor, we get the solid line  shown in
Fig.~\ref{FIG:Cracked} to illustrate the excellent agreement
between experiment and the TLL model. Our precise understanding of the spin Hamiltonian for DIMPY allows
us to make a direct comparison of our experimental findings to
numerical calculations. Using the field dependence of the
Luttinger parameter previously calculated for DIMPY using DMRG
~[\onlinecite{DIMPYthermodynamics}], for $H=9$~T we get $K=1.20$ or,
equivalently, $\gamma=-0.58$. This numerically obtained scaling exponent
is, to within experimental error, indistinguishable from our {\it ad hoc} experimental value, and the data collapse is almost as good. Equation (\ref{EQ:Sfactor}) can now be used to model the entire energy- and wavevector-dependent scattering intensity measured in our experiments. Numerically convoluting the analytical expression with the calculated spectrometer resolution, we get constant-energy cuts shown in solid lines in Figs. \ref{FIG:LowT}(b) and \ref{FIG:HighT}(b). Note that the only adjustable parameter in this fit is a single overall scaling factor for all temperatures and energies. Considering the experimental noise and systematic errors due to background subtraction, the level of agreement with experiment is excellent.

In summary, we show that the local dynamic structure factor for a
magnetized strong-leg quantum spin ladder has a finite-temperature
scaling form expected for a quantum-critical system. The
experimentally determined scaling exponent and scaling function are
in excellent agreement with those for an {\it attractive}
Tomonaga-Luttinger liquid. In fact, the results are fully consistent
with numerical calculations for the particular model spin
Hamiltonian of the target compound. We would like to emphasize that
neutron scattering experiments with energy resolution, intensity, and
signal-to-background ratios necessary for such studies were enabled
only very recently by major breakthroughs in neutron instrumentation
at user facilities such as ISIS.

This work was supported by the Swiss National Science Foundation,
Division 2. We would like to thank S.~M\"{u}hlbauer (FRM II,
Technische Universit\"{a}t M\"{u}nchen) for his involvement in
the early stages of this project. A.~Z. would like to thank Professor T.~ Giamarchi and Professor F.~H.~L.~Essler for numerous enlightening discussions on the subject of critical dynamics.

\bibliography{Dimpy_PRB}

\end{document}